\documentclass[10pt]{article}
\usepackage{opex3}
\usepackage{graphicx}
\usepackage{amsmath,amssymb}
\usepackage{bm}
\usepackage{subfigure}
\usepackage{cite}

\begin{document}
\title{Absorption enhancing proximity effects in aperiodic nanowire arrays}

\author{Bj\"{o}rn C. P. Sturmberg,$^{1\ast}$ Kokou B. Dossou,$^2$ Lindsay C. Botten,$^2$ Ara A. Asatryan,$^2$ Christopher G. Poulton,$^2$ Ross C. McPhedran$^1$, and C. Martijn de Sterke$^1$}
\address{$^1$ CUDOS and IPOS, School of Physics, University of Sydney, 2006, Australia
\\ $^2$ CUDOS, School of Mathematical Sciences, UTS, Sydney, 2007, Australia}
\email{$^{\ast}$b.sturmberg@physics.usyd.edu.au}

\begin{abstract}
Aperiodic Nanowire (NW) arrays have higher absorption than equivalent periodic arrays, making them of interest for photovoltaic applications. An inevitable property of aperiodic arrays is the clustering of some NWs into closer proximity than in the equivalent periodic array. We focus on the modes of such clusters and show that the reduced symmetry associated with cluster formation allows external coupling into modes which are dark in periodic arrays, thus increasing absorption. To exploit such modes fully, arrays must include tightly clustered NWs that are unlikely to arise from fabrication variations but must be created intentionally.
\vspace{3mm}
\end{abstract}


\bibliographystyle{osajnl}


\noindent

%
%
\section{Introduction}
%
%

Well-designed nanostructures are known to strongly enhance the absorption of photovoltaic devices \cite{Catchpole2006,Tsakalakos2008,Muskens2008,Leung2012,Atwater2010}, thereby allowing for the reduction of the active layer thickness to only a few micrometers. This significantly reduces costs, as less material is used and the material quality requirements are eased by shorter charge extraction distances.
Nanowire (NW) arrays are a prominent class of nanostructures that build on these advantages by incorporating radial p-n junctions, which further reduce the charge carrier diffusion length requirements. Periodic NW arrays have been studied theoretically \cite{Lin2009, Sturmberg2011} and numerous cost-effective fabrication methods are under development \cite{Seo2013,Turner-Evans2013,Wallentin2013,Garnett2011,Lu2012,Gunawan2010}.

It is known that the absorption of NW arrays is increased when they are arranged in aperiodic lattices, which may occur through random fabrication variations or may be designed on purpose \cite{Bao2010,Du2011a,Lin2011}.
Studying such aperiodic arrays is difficult due to the large parameter space and because of the broadband nature of the photovoltaic conversion process.
The origin of the absorption enhancement of aperiodic arrays has therefore remained unclear.

Here we avoid the difficulties of large parameter spaces by focussing on a common underlying property of aperiodic arrays: the localised clustering of NWs with reduced gaps between adjacent NWs.
This approach provides physical insights that apply to all aperiodic arrays; disordered or designed.
The importance of clustered NWs was highlighted by Lin and Povinelli \cite{Lin2011}, who showed strong field concentrations within NW clusters at peak absorption wavelengths.

\begin{figure}
\begin{center}
   \includegraphics[width=0.8\linewidth]{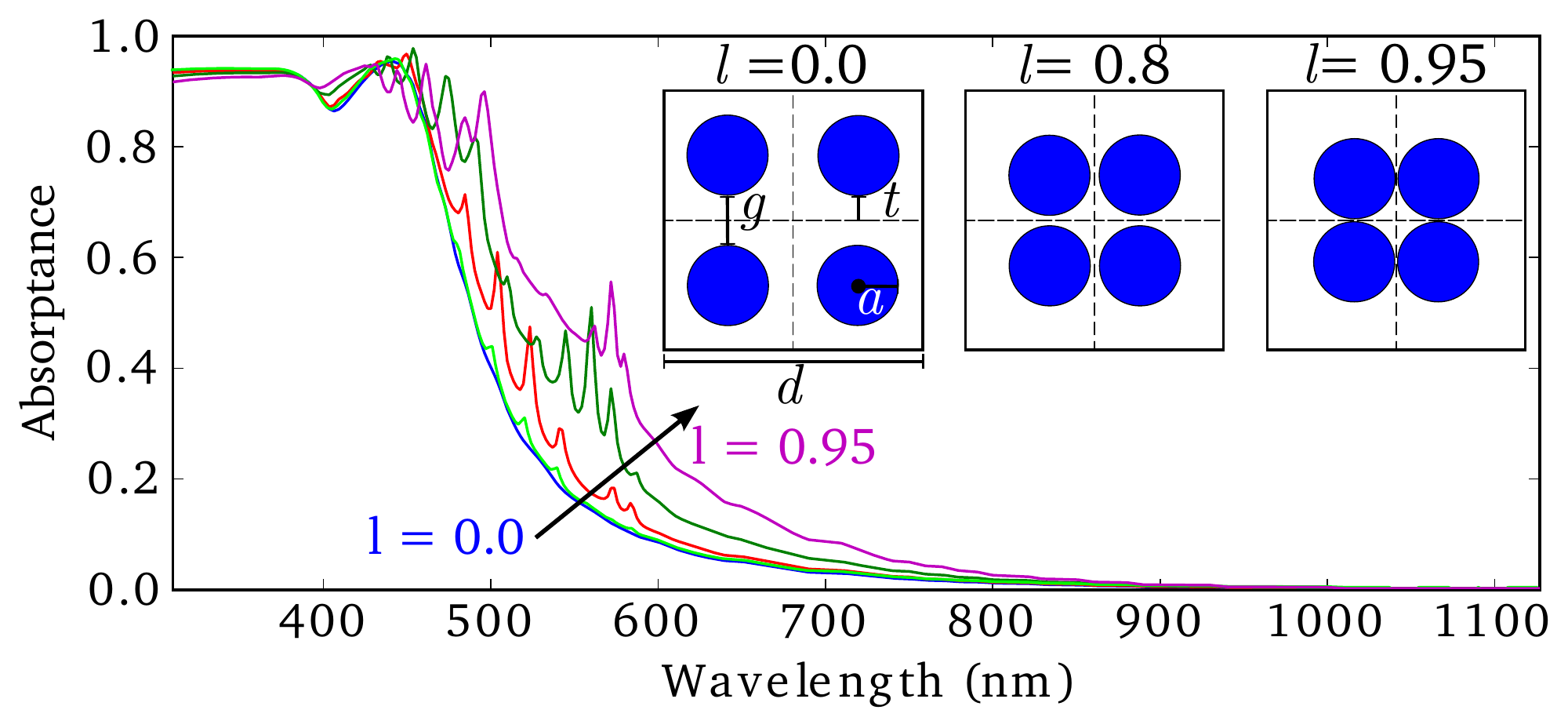}
\end{center}
\vspace{-7mm}
\caption{Absorption spectra of NW arrays with $a=31~{\rm nm}$, $f=30\%$, with increasing clustering $l = 0.0, 0.2, 0.5, 0.8, 0.95$. Inset shows cluster geometry where $a$ is the NW radius, $d$ is the unit cell dimension, $g$ is the gap size between NW surfaces, and $t$ is the distance a NW can be moved before touching its neighbour.}
\label{cluster_abs-4c}
\end{figure}


%
%
\section{Absorption Enhancement}
\label{Highabs}
%
%
We begin our study with clusters of 4 NWs of radius $a$, as shown in the inset of Fig.~\ref{cluster_abs-4c}. The NWs are initially periodically spaced in a square lattice with period $d/2$, in a unit cell of width $d$, before being gradually shifted towards the cell's centre. As soon as the NWs are moved, the periodicity of the structure becomes $d$.
We measure the {\it clustering level} $l$, as a fraction of the maximum distance the NWs can be moved before touching at $t = (d/4 - a)$. The {\it gap} between the NW surfaces is therefore $g = (1-l)\times(d/2 - 2a)$ (see Fig.~\ref{cluster_abs-4c} inset). Throughout the clustering process the volume fraction of the array is constant, $f = \pi a^2/(d/2)^2$.
Our simulations are carried out using a combination of the Finite Element Method (FEM) and the scattering matrix method as described earlier \cite{Dossou2012}. This calculation in the Bloch mode basis is numerically efficient and provides convenient access to physical quantities that are inaccessible in other methods, thereby allowing for greater insights.
Throughout this work the absorption spectra are calculated for normally incidence radiation from the short wavelength edge of the solar spectrum at $310~{\rm nm}$ to the band-edge of silicon at $1127~{\rm nm}$.


In Fig.~\ref{cluster_abs-4c} we show the absorption spectra for NW arrays with radius $a=31~{\rm nm}$ in a unit cell of $d=200~{\rm nm}$ ($f = 30\%$ silicon) as they are clustered with increasing $l$ from $0.0$ to $0.95$.
Note that the absorption increases with increasing clustering level of the arrays, particularly across the $\lambda = 500-600~{\rm nm}$ range.
The parameters of Fig.~\ref{cluster_abs-4c} are chosen to best illustrate the effect of clustering with clear spectra, rather than to achieve high photovoltaic efficiencies. Nonetheless, the increase in total absorption is significant.
At $\lambda=550~{\rm nm}$ the absorptance increases from $0.16$ in the unclustered array to $0.46$ at $l=0.95$.
The numerous sharp absorption peaks in this range indicate that resonant modes are excited in the clusters that are not present in the equivalent unclustered array (blue curve).

\begin{figure}
\begin{center}
   \includegraphics[width=0.99\linewidth]{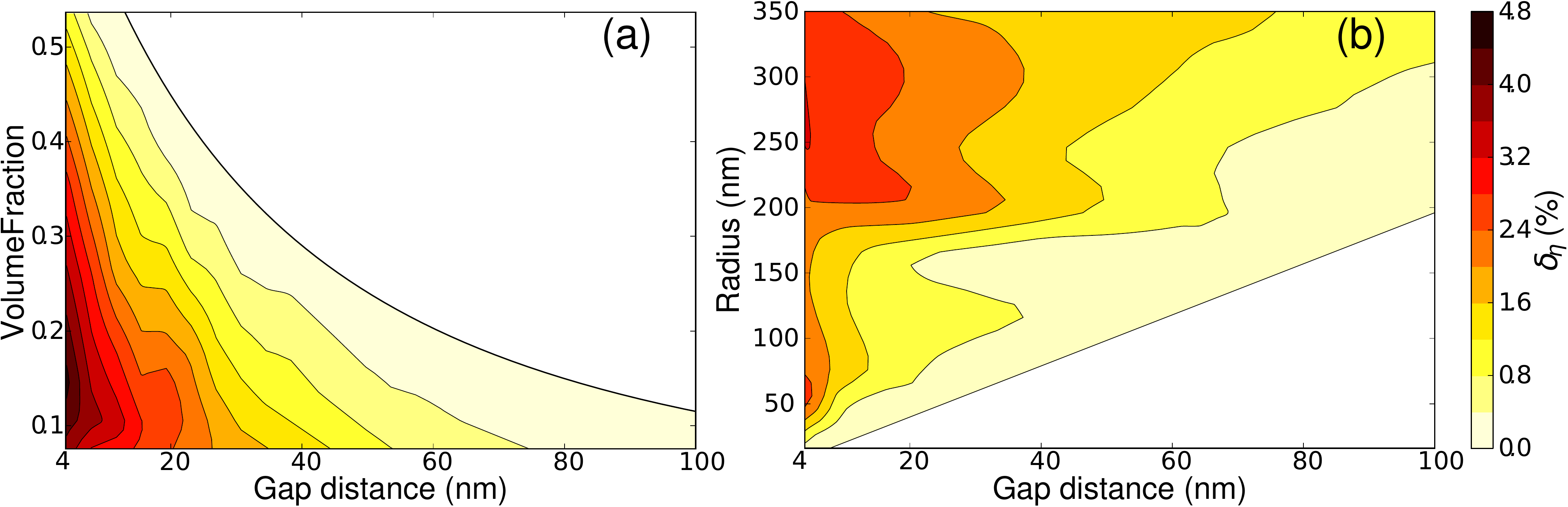}
\end{center}
\vspace{-5mm}
\caption{Ultimate efficiency increase $\delta\eta$ versus the gap between NW surfaces. (a) Arrays of different volume fractions with fixed NW radius $a=31~{\rm nm}$. (b) Arrays with different NW radii but fixed volume fraction, $f = 50\%$.
}
\label{fanfig}
\end{figure}

We quantify the efficiency enhancement produced through the ultimate efficiency, $\eta$ \cite{Shockley1961}, which weights the absorption spectrum by the solar spectrum and by the maximum fraction of energy which can be harvested.
In Fig.~\ref{fanfig} we show the ultimate efficiency enhancement $\delta\eta =\eta_{\rm clustered}~-~\eta_{\rm unclustered}$ for two intersecting slices through the clustered array parameter-space of $a, f, l$.
Figure~\ref{fanfig}(a) shows $\delta\eta$ for arrays with fixed radius arranged with a range of $f$, while in Fig.~3(b) the radii are varied and $f=50\%$ throughout.
This volume fraction produces near-optimal efficiencies when large radii are used, as in Fig.~\ref{fanfig}(b), and is consistent with multiple studies \cite{Lin2009,Lin2011,Sturmberg2012}.
When $a=239$ and $l=0.95$ in Fig.~\ref{fanfig}(b) the efficiency is $\eta = 22.54\%$.

In Figs.~\ref{fanfig}(a),(b) the efficiency monotonically increases with decreasing gap size, moving right to left, but substantial enhancements only occur once the NWs are tightly clustered, typically to a gap distance of less than $20\%$ of the NW radius.
Calculations with different volume fractions and radii, not shown here, show that these observations are true in general; accordingly all NW arrays benefit from increased efficiency with increased clustering, while substantial enhancements are only achieved for small gaps. The enhancement in the ultimate efficiency $\eta$ can be almost $5\%$ for the parameters in Fig.~\ref{fanfig}, a relative improvement of $85\%$.

%
%
\section{Excitation of Cluster Modes}
\label{origins}
%
%


\begin{figure}
\begin{center}
   \includegraphics[width=0.69\linewidth]{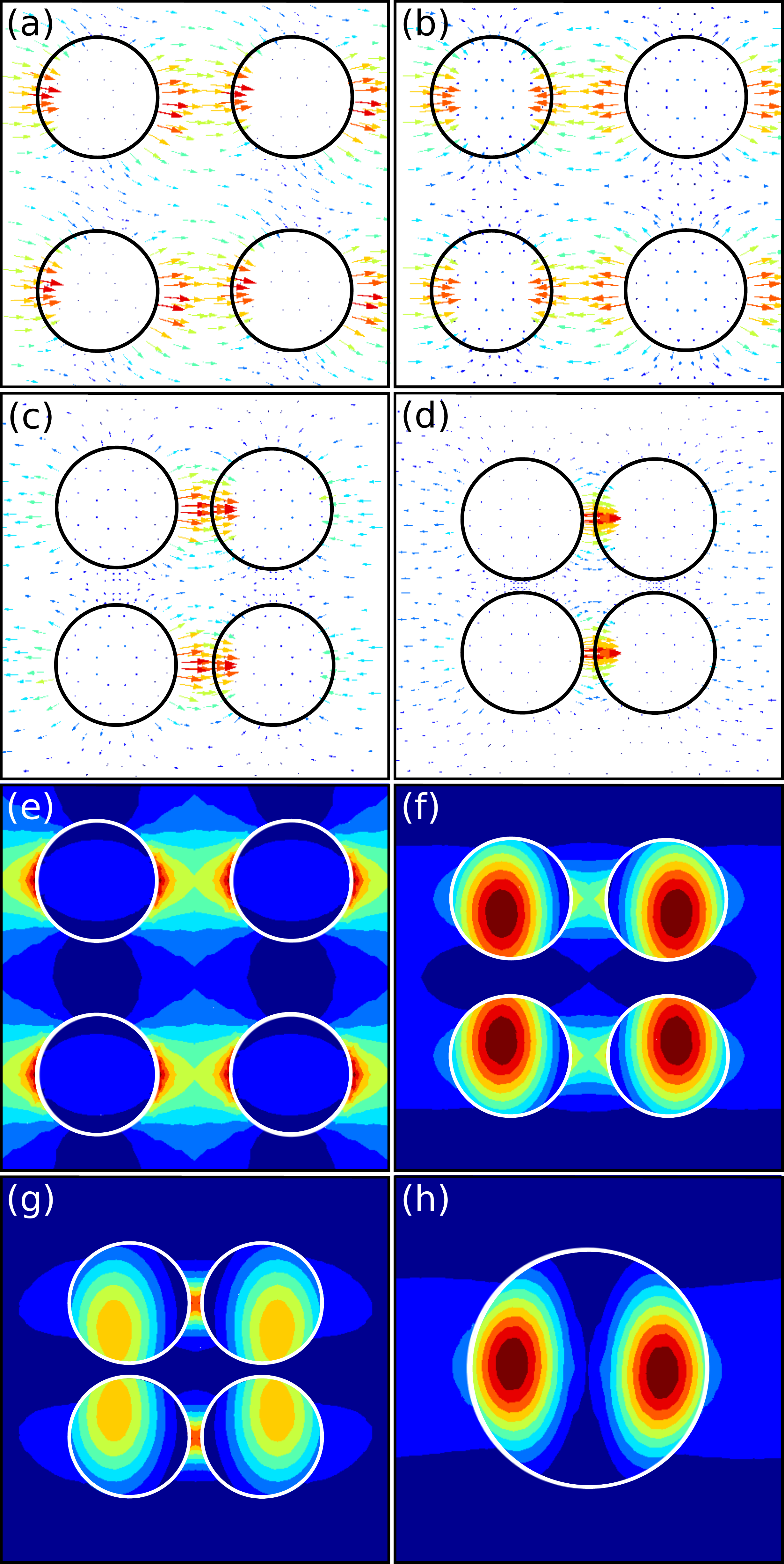}
\end{center}
\vspace{-3mm}
\caption{(a)--(d) Electric field vector of Bloch modes where colour and length indicate the field strength at the arrows' origin.
(a) Fundamental mode of the unclustered array;
(b) CKM of unclustered array;
(c) CKM with $l=0.5$;
(d) CKM with $l=0.8$.
(e)--(h) Bloch mode energy $\text{Re}(\epsilon)|\text{E}|^2$ where red and blue indicates high and low energy density, respectively.
(e) Fundamental mode of the unclustered array;
(f) CKM with $l=0.5$;
(g) CKM with $l=0.8$;
(h) KM of an array with twice the radius, {\sl i.e.}, $a=62~\text{nm}$.
For all figures $\lambda=550~\text{nm}$, $d=200~{\rm nm}$ and in (a)--(g) $a=31~\text{nm}$.
}
\label{fields-c}
\end{figure}


To understand the increased absorption we study the modal properties of periodic and clustered arrays. Though both support a multitude of Bloch modes, we need only consider those modes that couple to incident plane waves. We refer to such modes as bright modes, in contrast to dark modes, which cannot couple to external radiation at normal incidence. In our scattering matrix method the coupling coefficient between two modes (plane waves or Bloch modes) is derived from the overlap integral of the modes fields across the unit cell (see Eqs.~(64,~65) of \cite{Dossou2012}). From this we calculate the element of the transmission matrix for these modes as in Eq.~(69) in \cite{Dossou2012}, giving the coupling coefficient.
Calculating the coupling at the peak absorption wavelength of $550~{\rm nm}$, we find that unclustered arrays support a single bright mode, while clustered arrays have a second such mode.

To explain the existence of the second bright mode we show Bloch mode electric field vectors in Fig.~\ref{fields-c}(a)--(d). Here both colour and length indicate the field strength at the arrows' origin.
For the unclustered array we show the fundamental bright mode in Fig.~\ref{fields-c}(a) and a dark mode in Fig.~\ref{fields-c}(b).
The bright mode has all strong field components directed from left to right, while the dark mode has field components of equal strength oriented in opposite directions.
The overlap integral with a plane wave therefore vanishes for the dark mode of Fig.~\ref{fields-c}(b). In contrast it is non-zero for the bright mode of Fig.~\ref{fields-c}(a) for horizontally polarised incoming light.
Vertically polarized plane waves couple to a vertically oriented mode degenerate with that in Fig.~\ref{fields-c}(a).

\begin{figure}
\begin{center}
   \includegraphics[width=0.6\linewidth]{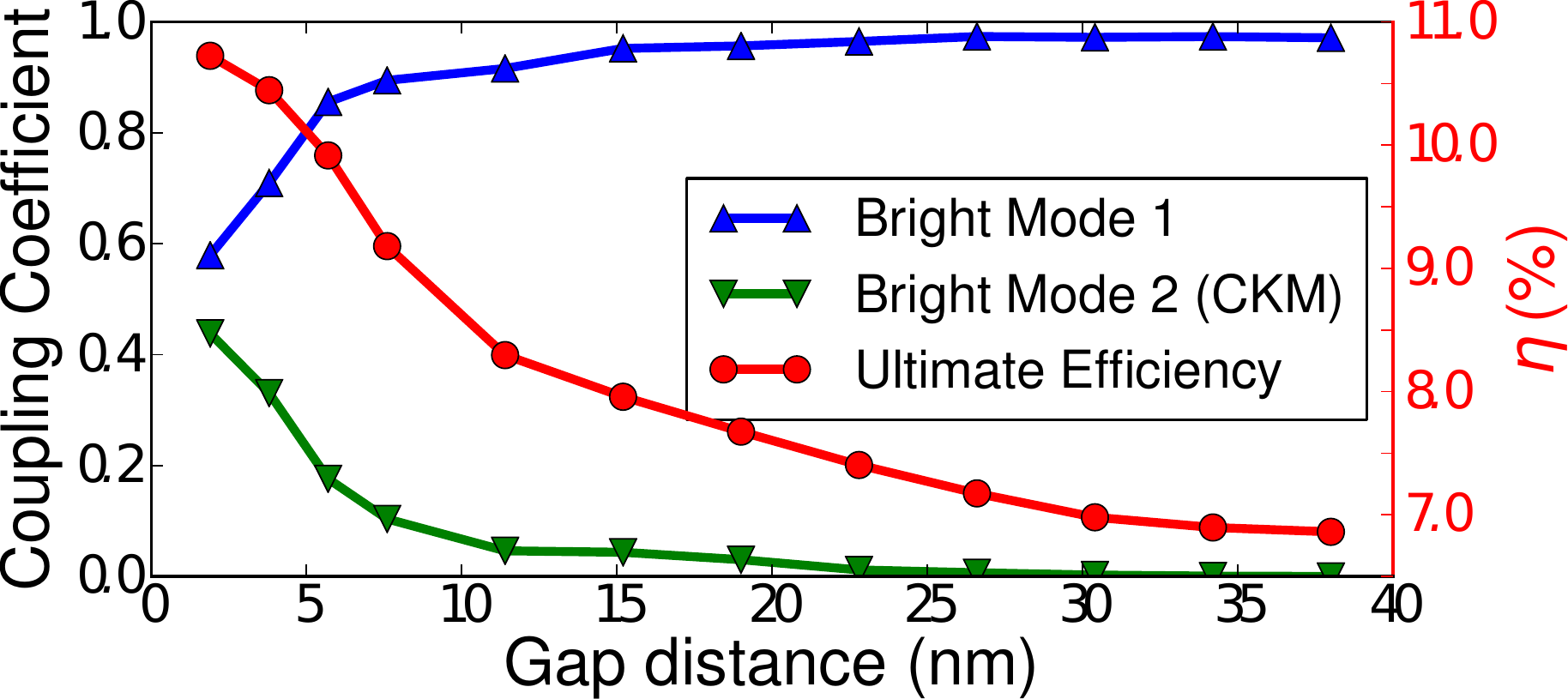}
\end{center}
\vspace{-5mm}
\caption{Coupling coefficient (left axis) of the incident plane wave to the fundamental mode (blue upward triangles) and the second bright mode (green downward triangles) versus gap between NW surfaces. The ultimate efficiency $\eta$ of the clusters is also shown (red circles and right axis).
}
\label{coupling}
\end{figure}

Once the NWs are clustered, the symmetry of the dark mode of Fig.~\ref{fields-c}(b) is lowered as the fields between the cylinders strengthen, and those on the outside of the cluster weaken.
As shown in Figs.~\ref{fields-c}(c) and (d), this results in an electric field that is predominantly oriented from left to right, allowing for coupling to an incoming field. The asymmetry strengthens with increasing clustering, leading to stronger coupling.
This is shown in Fig.~\ref{coupling} where the coupling coefficient of the incident plane wave to the fundamental mode at $\lambda = 550~{\rm nm}$ decreases with decreasing gap (blue) as the coupling into the second bright mode increases (green).


To be highly absorbing a Bloch mode must not only couple to the incident field but also concentrate its energy within the absorbing material \cite{Sturmberg2011}. In Figs.~\ref{fields-c}(e)--(h) we show the energy concentration $\text{Re}(\epsilon)|\text{E}|^2$ of modes at $\lambda = 550~{\rm nm}$. At this wavelength, the bright mode of the unclustered array in Fig.~\ref{fields-c}(e) has little energy within the silicon, resulting in the low absorption of this structure (see Fig.~\ref{cluster_abs-4c}).
In contrast, the bright mode of the clustered arrays has significant energy in the silicon, as shown in Fig.~\ref{fields-c}(f) and (g) for $l=0.5$ and $0.8$, respectively. When combined with strong coupling this produces the enhanced absorption of Fig.~\ref{cluster_abs-4c}.
Consistent with this, the red curve in Fig.~\ref{coupling} shows that the clusters' ultimate efficiency improves as the coupling into the second bright mode increases. Though the coupling is shown for a single wavelength, this is a broadband effect which occurs for all wavelengths below the bright mode's cut-off at $\lambda = 600$~nm. The absorption of this mode, which is associated with all absorption peaks in Fig.~\ref{cluster_abs-4c}, is enhanced by its low group velocity and strong Fabry-P\'{e}rot resonances between the top and bottom interfaces. Both of these effects are characteristics of the {\it Key Modes} (KMs) of periodic NW arrays \cite{Sturmberg2011}. We therefore refer to the bright mode of the cluster as a {\it Clustered Key Mode} (CKM). In Fig.~\ref{fields-c}(h) we show how the energy concentration of the CKM is similar to the KM of a periodic array with doubled radius NWs ($a=62~{\rm nm}$).

%
%
\section{Discussion and Conclusion}
%
%
We have concentrated on one consequence of aperiodicity, namely that, compared to periodic arrays of the same density, it inevitably leads to the clustering of NWs. The associated symmetry lowering leads to the coupling of plane waves into a previous inaccessible mode, thereby significantly increasing the array's absorption. When the clustering is quite pronounced these Clustered Key Modes are similar to the strongly absorbing Key Modes that exist in periodic arrays of large radius NWs. Though we concentrated on the clustering of 4 NWs, we also studied clusters of 2, 3, and 5 NWs and came to similar conclusions. This suggests that our results are general and not specific to the geometry studied here.

The absorption enhancement mechanism identified here differs from that in structures consisting of NWs of different radii. There, the diversity of the radii produces a wide spectrum of modal absorption resonances across the solar spectrum, hence increasing the absorption efficiency \cite{Sturmberg2012}. Here, in contrast, the lowering of the translational symmetry through the clustered NW arrangement produces bright modes which are inaccessible in unclustered arrays.

Our results in Fig.~\ref{fanfig}, and others not shown here, show that, compared to the unclustered periodic case with period $d/2$, clustering always increases absorption. The degree to which it does so depends not only on the cluster parameters but also on the array's global properties such as the value of $d$. It is therefore difficult to draw specific quantitative conclusions. However, we have observed that the clustering only significantly enhances the absorption when the NW are brought close, typically such that the gap between their surfaces is approximately 20\% of the NW radius. This is consistent with Fig.~4(b) of Lin and Povinelli \cite{Lin2011}.
As a consequence of this, the random variations in NW positioning that arise naturally during fabrication of periodic arrays are unlikely to be sufficient to exploit the efficiency enhancement potential of aperiodic arrays. Rather, the potential of aperiodic arrays may only be harnessed by purpose fabrication of aperiodic arrays with closely clustered NWs.

%
%
\section*{Acknowledgments}
%
%
We acknowledge helpful discussions with Dr Kylie Catchpole and Dr Thomas White.
This work was supported by the Australian Renewable Energy Agency, and by the Australian Research Council (ARC) Center of Excellence for Ultrahigh bandwidth Devices for Optical Systems (Project CE110001018) and by an ARC Discovery Grant. Computation resources were provided by the National Computational Infrastructure, Australia.

%
%
%
%

\end{document}